\documentclass[sigconf]{acmart}

\def\BibTeX{{\rm B\kern-.05em{\sc i\kern-.025em b}\kern-.08emT\kern-.1667em\lower.7ex\hbox{E}\kern-.125emX}}

\copyrightyear{2019}

\acmConference[]{}{}{}

\usepackage{hyperref}
\usepackage{lineno}
\usepackage{textcomp}
\usepackage{listings,multicol}
\usepackage{multirow}
\usepackage[framemethod=tikz]{mdframed}
\usepackage[textsize=tiny]{todonotes}
\usepackage{colortbl}

\lstdefinelanguage{nmodl}{
  morekeywords={%
    NEURON, SUFFIX, USEION, READ, WRITE, RANGE, UNITS, PARAMETER, TITLE,
    ASSIGNED, STATE, BREAKPOINT, SOLVE, METHOD, cnexp, derivimplicit, sparse, INITIAL, NRN_STATE,
    DERIVATIVE, KINETIC, FUNCTION, LOCAL, UNITSOFF, UNITSON, ENDVERBATIM, VERBATIM, PROCEDURE
  },
  morecomment=[l]{\:},
}

\begin{document}

%
\title{An optimizing multi-platform source-to-source compiler framework for the NEURON MODeling Language}

%
\author{Pramod Kumbhar}
\affiliation{%
  \institution{Ecole Polytechnique F\'ed\'erale de Lausanne (EPFL)}
  \streetaddress{}
  \city{}
  \country{}}
\email{}

\author{Omar Awile}
\affiliation{%
  \institution{Ecole Polytechnique F\'ed\'erale de Lausanne (EPFL)}
  \streetaddress{}
  \city{}
  \country{}}
\email{}

\author{Liam Keegan}
\affiliation{%
  \institution{Ecole Polytechnique F\'ed\'erale de Lausanne (EPFL)}
  \streetaddress{}
  \city{}
  \country{}}
\email{}

\author{Jorge Blanco Alonso}
\affiliation{%
  \institution{Ecole Polytechnique F\'ed\'erale de Lausanne (EPFL)}
  \streetaddress{}
  \city{}
  \country{}}
\email{}

\author{James King}
\affiliation{%
  \institution{Ecole Polytechnique F\'ed\'erale de Lausanne (EPFL)}
  \streetaddress{}
  \city{}
  \country{}}
\email{}

\author{Michael Hines}
\affiliation{%
  \institution{Yale University}
  \streetaddress{}
  \city{}
  \country{}}
\email{}

\author{Felix Sch\"urmann}
\email{felix.schuermann@epfl.ch}
\affiliation{
  \institution{Ecole Polytechnique F\'ed\'erale de Lausanne (EPFL)}
  \streetaddress{}
  \city{}
  \country{}}

%
\renewcommand{\shortauthors}{Kumbhar P., et al.}

\begin{abstract}

Domain-specific languages (DSLs) play an increasingly important role in the generation of high
performing software. They allow the user to exploit specific knowledge encoded in the constructs for the generation of code
adapted to a particular hardware architecture; at the same time, they make it easier to generate optimized code for
a multitude of platforms as the transformation has to be encoded only once. Here, we describe a new code generation
framework (NMODL) for an existing DSL in the NEURON framework, a widely used software for massively parallel
simulation of biophysically detailed brain tissue models. Existing NMODL DSL transpilers lack either essential features 
to generate optimized code or capability to parse the diversity of existing models in the user community. Our NMODL 
framework has been tested against a large number of previously published user models and offers high-level domain-specific
optimizations and symbolic algebraic simplifications before target code generation. Furthermore, rich analysis tools
are provided allowing the scientist to introspect models. NMODL implements multiple SIMD and SPMD targets optimized
for modern hardware. Benchmarks were performed on Intel Skylake, Intel KNL and AMD Naples platforms. When
comparing NMODL-generated kernels with NEURON we observe a speedup of up to 20x, resulting into overall speedups
of two different production simulations by $\sim$10x. When compared to a previously published SIMD optimized version that 
heavily relied on auto-vectorization by the compiler still a speedup of up to $\sim$2x is observed.
\end{abstract}

%
%

%

%
\maketitle

\section{Introduction}
\label{sec:introduction}

The use of large scale simulation in modern neuroscience is becoming increasingly important 
(e.g. ~\cite{Markram2015,Arkhipov2018a,Schmidt2017})
and has been enabled by substantial performance progress in neurosimulation engines over the last decade and a half 
(e.g. ~\cite{Migliore2006,Hines2011,Kumbhar2019,Morrison2005,Kunkel2014,Jordan2018a}).

While excellent scaling has been achieved on a variety of platforms with the conversion to Single
Instruction Multiple Data (SIMD)
implementations, domain specific knowledge expressed in the models is not yet optimally used.
In other fields, the use of DSLs and subsequent
code-to-code translation have been effective in generating high performing codes and allowing easy
adaptation to novel architectures (\cite{Starruss2014,Devito2011,Rathgeber2012,Schmitt2014,Membarth2012}).
This is becoming more important as the architectural diversity of computers is increasing as a reaction
to mutually exclusive design trade offs forced by more evident physical and economic limitations when going to
the next generation of computing systems.

Motivated by these observations, we have revisited the widely adopted NEURON simulator~\cite{Hines1997}, which enables 
simulations of biophysically detailed neuron models on computing platforms ranging from desktop to petascale 
supercomputers, and which has over 2,000 reported scientific studies using it.
NEURON supports networks of neurons 
with complex branched anatomy and biophysical properties such as multiple channel types, inhomogeneous channel distribution,
ionic accumulation and diffusion-reactions.
One of the key features of the NEURON simulator is
extendability via a domain specific language (DSL) layer called the
NEURON Model Description Language (NMODL)~\citep{Hines2000}.
NMODL allows model authors to extend NEURON by incorporating
a wide range of membrane and intracellular submodels such as voltage and ligand gated channels,
ionic accumulation and diffusion, and synapse models.
The domain scientists can easily express these channel properties in terms of algebraic
and ordinary differential equations, kinetic schemes, and finite state machines in NMODL.
Working at a descriptive level allows easy development of models while focusing on biological
aspects rather than worrying about lower level implementation details such as solver methods,
threads, memory layout, parallel execution etc.

\begin{figure}
  \includegraphics[width=0.5\textwidth]{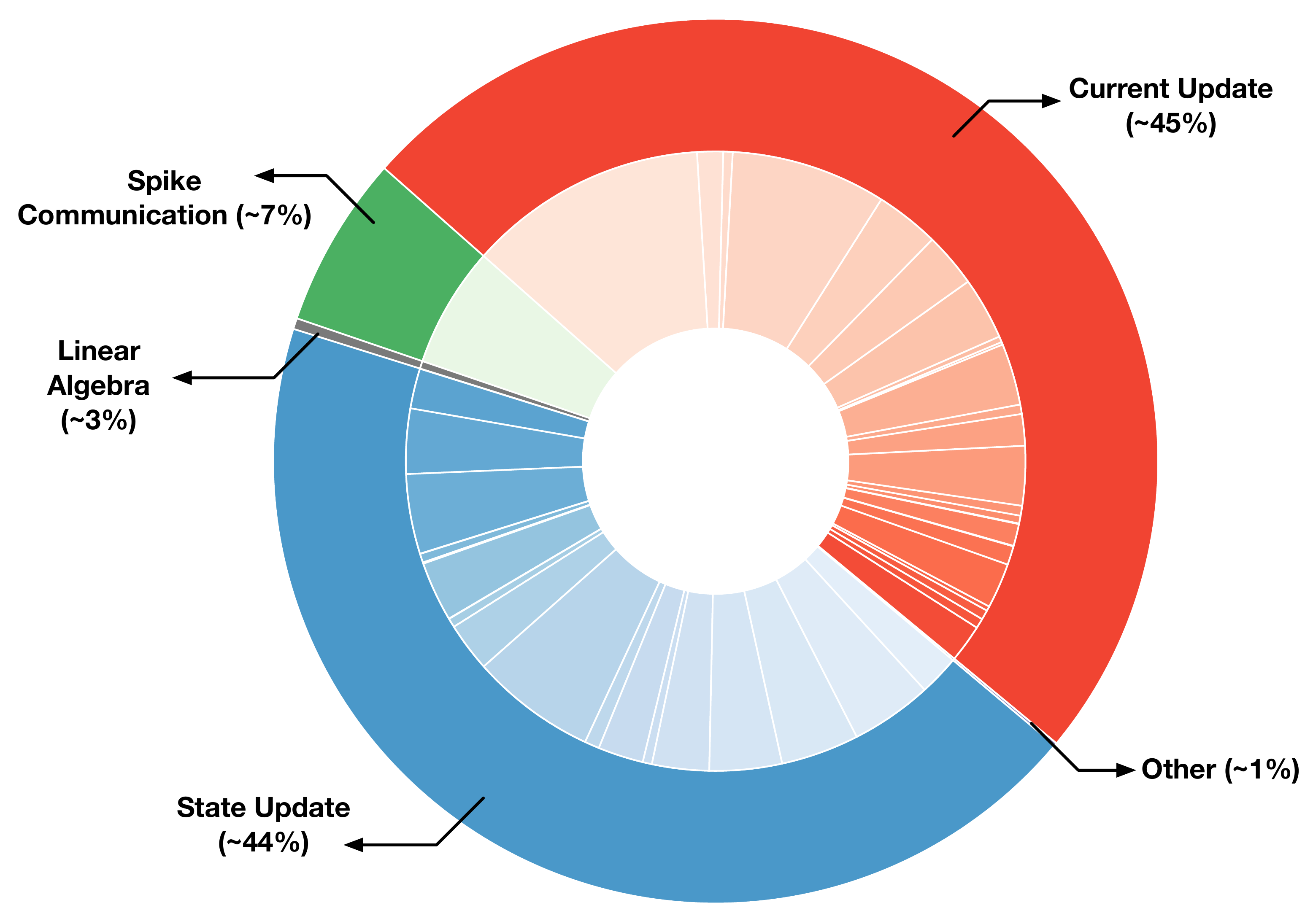}
  \caption{
  Execution profile of hippocampus CA1 model showing the percentage of
  time spent in various parts of the NEURON simulator. More than 90\%
  of the execution time is spent in the kernels generated from the NMODL DSL.}
  \label{fig:simulation-profile}
\end{figure}

The rate limiting aspect for performance of NEURON simulations is the execution of channels
and synapses written in the NMODL DSL. The code generated from NMODL often accounts
for more than \emph{80\%} of overall execution time.
For example, \autoref{fig:simulation-profile} shows the execution profile of a Hippocampus model~\cite{HumanBrainProject} with about 670 thousand compartments
and about 4.3 million synapses simulated for one second of biological time
(see \autoref{sec:benchmark} for details).
This model has 17 morpho-electrical types (me-types) where electrical firing behavior of an metype is governed by the distribution of various channel types across the morphological cell body. Each channel is described in a separate NMODL file
and typically contains two kernels : \emph{state update} and \emph{current update}.
The outer ring of the pie chart shows the percentage of time spent
in the different parts of the NEURON simulator.
The \emph{current update} and \emph{state update} from the NMODL DSL accounts
for more than \emph{90\%} of the simulation time.
The execution time of individual mechanisms are shown by arcs in the inner circle.
Even in a strong scaling scenario, it has been shown that the spike communication
remains small and channel/synapse computations dominate the overall execution time~\cite{Ovcharenko2015}.

There are more than six thousand NMODL files that are shared by the NEURON user community
on the ModelDB platform~\cite{Nadkarni1996}.
As the type and number of mechanisms differ from model to model,
hand-tuning of the generated code is not feasible.
The goal of our NMODL Framework is to provide a tool that can parse all existing models,
and generate optimized code from NMODL DSL code, which is responsible for more than \emph{80\%} of the total simulation time.

\section{Related Work}
\label{sec:relatedwork}

The reference implementation for the NMODL DSL specification is found in
\emph{nocmodl}~\cite{Hines1989}, which is a component in the NEURON simulator.
Over the years \emph{nocmodl} underwent several iterations of development and
gained support for a number of newer language constructs and a \emph{C} code
generator backend for the NEURON simulator. \emph{Nocmodl} uses
\emph{lex/flex}~\cite{Paxson1987} and \emph{yacc/bison}~\cite{Corbett1985} for
its scanner and parser implementation respectively. One of the major limitations
of \emph{nocmodl} is its lack of flexibility. Instead of constructing an
intermediate representation, such as an \emph{Abstract Syntax Tree} (AST), it
performs many code generation steps on the fly, while parsing input as part of
the \emph{bison} production rules. This leaves little flexibility for performing
global analysis, optimizations, adapting code generation for different languages
or targeting a different simulator altogether. The CoreNEURON library uses a
modified version of \emph{nocmodl} called \emph{mod2c}~\cite{BlueBrainProject2015}.
\emph{Mod2c} duplicates most of the
legacy code with the modifications needed to generate code with new data
structures present in CoreNEURON, different memory layouts and GPU support based
on the OpenACC programming model.  But, as \emph{mod2c} shares most of its
implementation with \emph{nocmodl}, it still has some of the same limitations as
\emph{nocmodl}.
\emph{Pynmodl}~\cite{Marin2018} is a Python based parsing and post-processing
tool for NMODL. The primary focus of \emph{pynmodl} is to parse and translate
NMODL DSL to other computational neuroscience DSLs such as
LEMS~\cite{Cannon2014}.
\emph{Pynmodl} supports a subset of the
NMODL DSL specification and does not support lower level code generation for a
particular simulator. The \emph{modcc} source-to-source compiler is being
developed as part of the Arbor simulator~\cite{Akar2019}. It is able to generate
from NMODL DSL code, optimized C++/CUDA to be used with the Arbor simulator. Its
lexer and parser have been hand-implemented in C++ and generate an intermediate
AST representation. However, it only implements a subset of the NMODL DSL
specification and hence is only able to process a modest number of existing
models available in ModelDB~\cite{Nadkarni1996}. For a more comprehensive review
of current code-generator techniques in computational
neuroscience we refer the reader to Blundell et al~\cite{Blundell2018}. In
summary, we conclude that current tools either lack support for the full NMODL DSL
specification, lack the necessary flexibility to be used as a generic code
generation framework, or are unable to adequately take advantage of modern hardware
architectures, and thus are missing out on available performance from modern
computing architectures.

\section{NMODL DSL}
\label{sec:nmodl_dsl}

\begin{figure*}
  \includegraphics[width=1.0\textwidth]{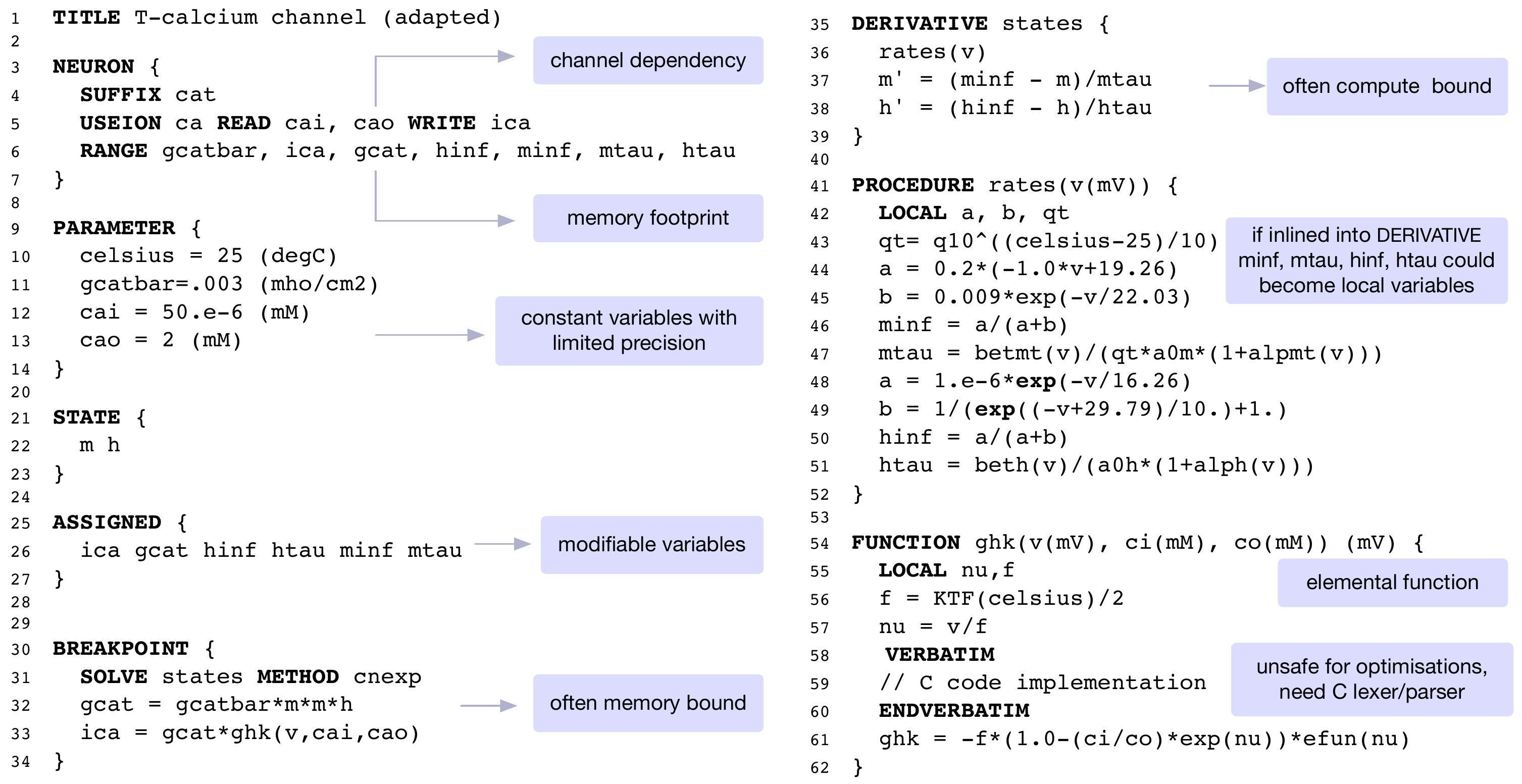}
  \caption{NMODL example of voltage-gated calcium channel showing different NMODL constructs and
           summary of optimization information available at DSL level}
  \label{fig:nmodl-example}
\end{figure*}

To understand important code generation aspects for improving performance, a simple example of a voltage-gated calcium channel written in NMODL DSL is presented in \autoref{fig:nmodl-example}. Due to space
constraints the example is modified to highlight important language constructs.

Each named block in the NMODL DSL has the general form of \texttt{KEYWORD \{ statements \}}. There
are more than 90 different constructs or keywords in the language. Keywords are printed in all uppercase
and marked with boldface in the example. The \texttt{NEURON} block specifies the name of the
mechanism, ion usage, and variables used in the model. The \texttt{SUFFIX} keyword identifies
this as a density mechanism (as opposed to \texttt{POINT\_PROCESS} which would instead identify it as a synapse or electrode class whose instances are localized to a single point on the neuron). The
\texttt{USEION} statement indicates the interaction of this mechanism with those other mechanisms at the same location sharing the same ion type. The \texttt{RANGE}
keyword indicates that the encompassed variables are functions of position (i.e. each of these variables can have
a different value in each compartment). \texttt{LOCAL} variables are block scoped variables in the block. Internal to the mod file, variables that are used as model parameters are defined
in the \texttt{PARAMETER} block.
The \texttt{ASSIGNED} block is used to declare variables that
can appear on the left hand side of assignment statements (i.e. modifiable variables). If a model uses
differential equations, algebraic equations, or kinetic reaction schemes, their dependent variables
are listed in the \texttt{STATE} block. The \texttt{BREAKPOINT} block is used to update current and
conductance at each time step. The \texttt{DERIVATIVE} block contains differential equations of the
form $y' = expr$ that are used to assign values to the derivatives of \texttt{STATE} variables. The
\texttt{SOLVE} statement is used to specify the integration scheme (see \autoref{sec:ode})
The \texttt{PROCEDURE} and \texttt{FUNCTION} represents callable functions where the only difference is that the
\texttt{FUNCTION} variation can return a value. Users can also use \texttt{VERBATIM} constructs to embed
\emph{C} code directly into the DSL. The enclosing statements will be copied to the generated code
in place. This offers flexibility to implement lower level functionality not exposed by NMODL DSL, but
also makes it difficult to generate portable code. More information about NMODL specification can be found
in~\cite{Hines2000}.

At the DSL level there is a lot of information implicitly expressed that can be used to generate
efficient code and expose more parallelism. Some of the examples are: 
\begin{itemize}
  \item \texttt{USEION} statement describes the dependency of this mechanism with other ion channels
   (e.g. \emph{ca}). This information can be used to build the runtime dependency graph to exploit
   micro-parallelism~\cite{Magalhaes2019}
  \item \texttt{PARAMETER} block describes the variables that are constant during runtime, often
   with limited precision requirements and often with small range of values (e.g. \emph{gcatbar}).
   This information can be used to improve memory access cost and reduce memory footprint.
  \item \texttt{ASSIGNED} statement describes modifiable variables. These variables can be allocated
   in fast memory to reduce access latency (e.g. \emph{minf}).
  \item \texttt{DERIVATIVE}, \texttt{KINETIC} and \texttt{SOLVE} describes ODEs which can be analyzed
   and solved analytically to improve the performance as well as accuracy.
  \item \texttt{BREAKPOINT} describes current and conductance update at each time step. If the current and
   voltage relation is ohmic then one can use analytical expression instead of numerical derivatives to
   improve the accuracy as well as performance.
  \item \texttt{PROCEDURE} does not allow a return value and hence often users
  use \texttt{RANGE} variables (e.g. \emph{minf, mtau}). If plotting of such variables is not required,
  procedures can be inlined at DSL level (e.g. \emph{rates}) to eliminate RANGE variables and thereby significantly
  reduce memory access cost as well as memory footprint.
\end{itemize}

To use this information and perform such optimizations, often global analysis of the NMODL DSL is
required. For example, to perform inlining of a \texttt{PROCEDURE} one needs to find all function
calls and recursively inline the function bodies. To verify if a variable can be made \emph{const}, one needs -to find all its usages..
As \emph{nocmodl} lacks the intermediate AST representation, this
type of analysis is difficult to perform and these optimizations are not implemented. The NMODL code
generation framework is designed to exploit all such information from DSL specification and perform
optimizations.

\begin{figure*}
  \includegraphics[width=1.0\textwidth]{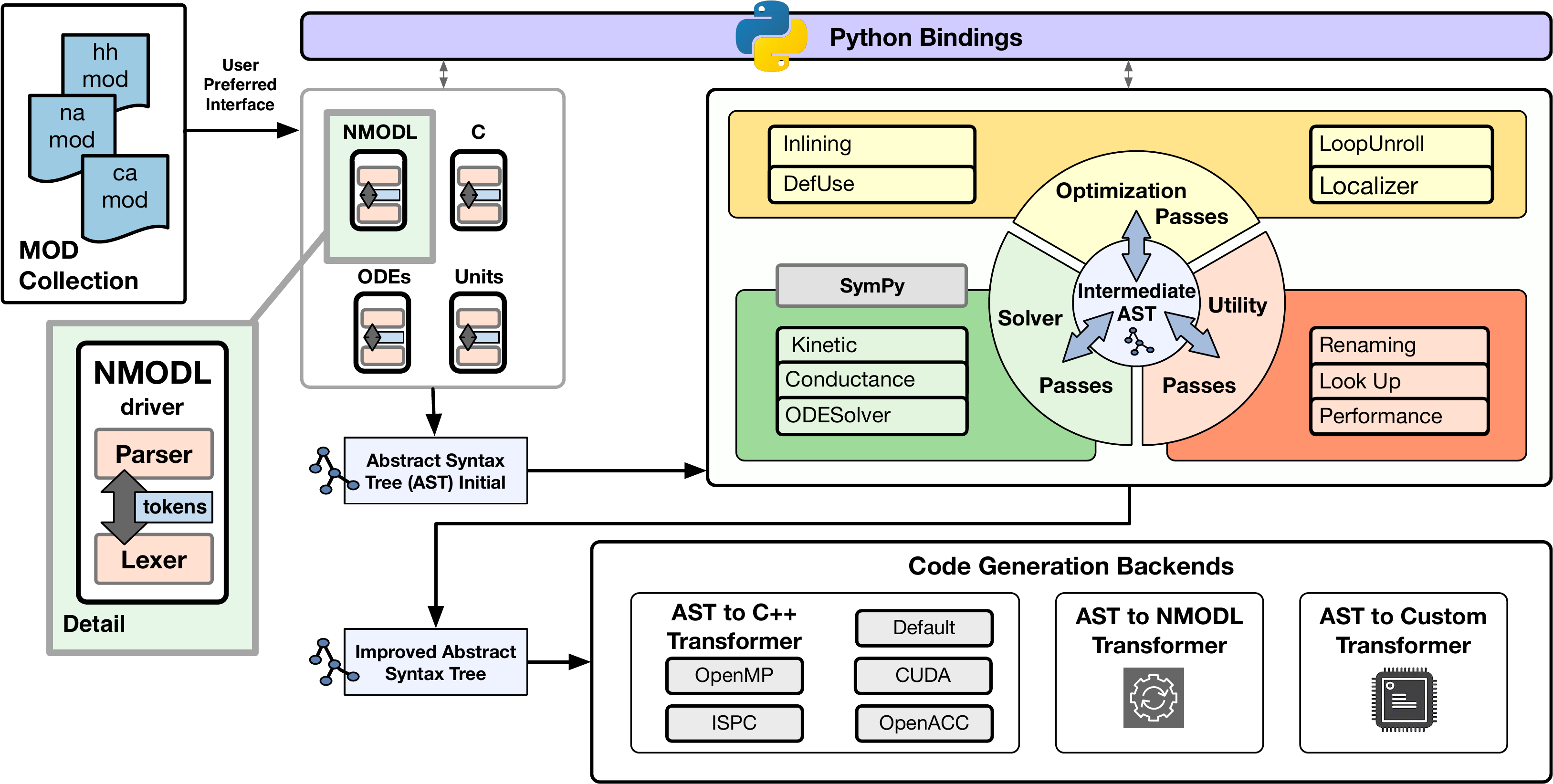}
  \caption{Architecture of the NMODL Code Generation Framework showing:  A) Input NMODL files are
  processed by different lexers \& parsers generating the AST; B) Different analysis and
optimisation passes further transform the AST; and C) The optimised AST is then converted to low
level C++ code or other custom backends}
  \label{fig:architecture}
\end{figure*}

\section{Design and Implementation}

The overall design of the NMODL Framework can be broken down into six components: lexer/parser
implementation, AST and DSL level optimisation passes, ODE solvers, Python bindings, and
finally code generation passes. \autoref{fig:architecture} summarizes the overall architecture.

\subsection{Lexer and Parser}
\label{subsec:lexer-parser}

As in any compilation process, the first two steps performed on an input NMODL DSL are \emph{lexing} and
\emph{parsing}. The lexer implementation is based on the popular \texttt{flex} package and
\texttt{bison} is used as the parser generator. To make the lexer and parser fully
reentrant~\cite{Pankevich2016}, we makes use of the C++ Bison Interface parser implementation.
The bison grammar is based on \emph{nocmodl}, which is a reference implementation for the NMODL DSL,
making our implementation fully compatible with the NMODL language specification. 
As parser rules are executed, appropriate C++ classes are instantiated to construct
an Abstract Syntax Tree (see ~\autoref{subsec:ast}). As opposed to
\emph{nocmodl} and \emph{mod2c} our approach
allows us to keep parsing and code generation steps completely separated and an arbitrary number
of intermediate analysis and optimization steps can be interposed. The ODEs, units and inline C
code (from \texttt{VERBATIM} blocks) need extra processing during further distinct compiler passes. To parse
these constructs, we have implemented separate lexers and parsers using \texttt{flex} and
\texttt{bison}.

\subsection{Abstract Syntax Tree Representation}
\label{subsec:ast}

\begin{figure}
  \includegraphics[width=0.5\textwidth]{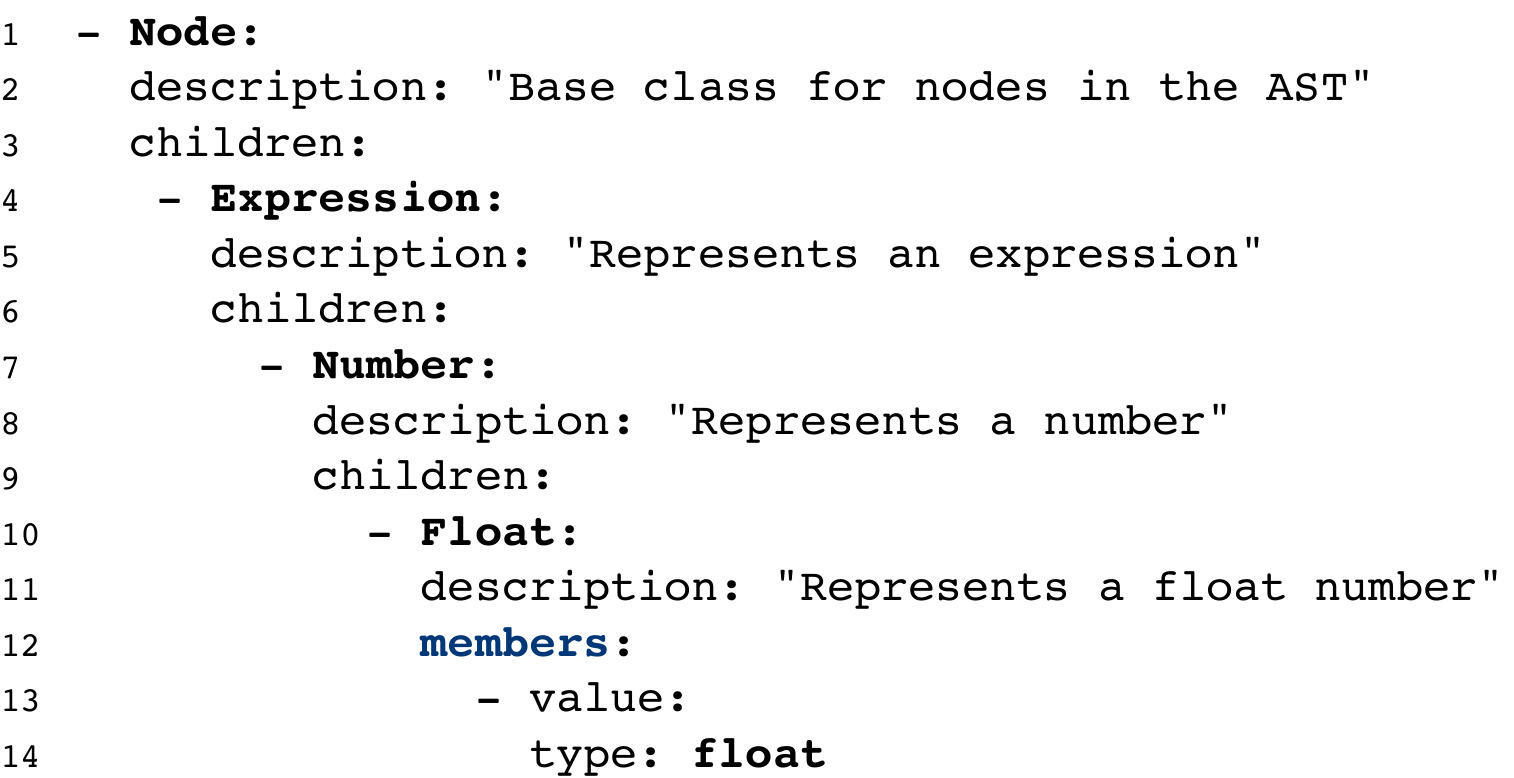}
  \caption{An example of YAML specification for AST base nodes like \emph{Node}, \emph{Expression} and child node \emph{Float}}
  \label{fig:ast-node}
\end{figure}

The goal of the NMODL Framework is to implement a general purpose parsing and code generation tool not
tied to a particular simulator. For this, all language semantics need to be preserved in the
Abstract Syntax Tree (AST)
representation so that different higher level tools can use this information for different purposes.
With more than 90 keywords, 130 different AST node types are required to represent all language
semantics. Each node of the AST is implemented as a \emph{C++} class and represents a syntactic
element, or a group of syntactic or semantic elements of the DSL. To avoid having to implement more
than 130 C++ classes by hand, we used the declarative approach for AST design using a YAML
specification~\cite{Ingerson2001}. \autoref{fig:ast-node} shows an excerpt of the YAML
specification of the DSL. This YAML specification is translated to
\emph{C++} classes using the Python jinja2 template engine~\cite{Lord2008}. Using this approach we are
able to generate most of the AST representation and AST visitor framework in a compact code
base. This approach provides flexibility to extend the AST design with minimal changes and hence
higher productivity.

\begin{figure}
  \includegraphics[width=0.5\textwidth]{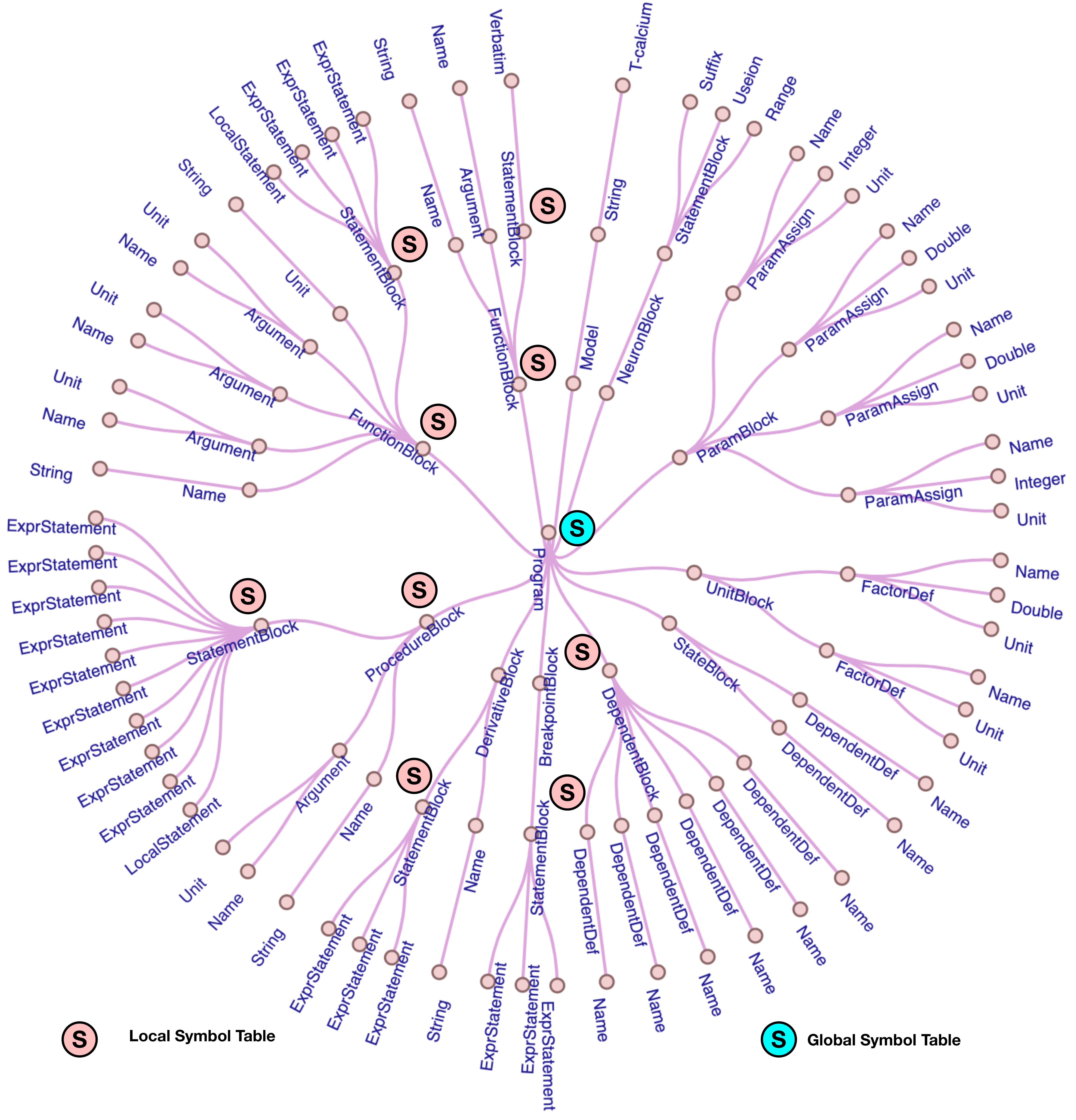}
  \caption{In memory AST representation of the NMODL code from ~\autoref{fig:nmodl-example} showing different AST node types and symbol tables created.}
  \label{fig:ast}
\end{figure}

The hierarchy in the YAML specification represents
inheritance relationships with C++ AST classes. The \emph{member} specification describes \emph{member}
variables with their types in \emph{C++} classes. 

\autoref{fig:ast} shows an example of the AST in-memory representation constructed by the NMODL
parser for the example discussed in \autoref{sec:nmodl_dsl}.
As the NMODL language supports block scopes, scoped symbol tables~\cite{Pivak2016} are created at each block scope and attached to the corresponding AST node.

\subsection{Optimization Passes}
\label{subsec:optimizationpasses}

Modern compilers implement various passes for code analysis, code transformation, and optimized code
generation. Optimizations such as constant folding~\cite{Wegman1991}, inlining~\cite{Das2003}, and loop
unrolling~\cite{Sarkar2001} are commonly found in all of today's major compilers. For example, the LLVM
compiler framework~\cite{Lattner:2004:LCF:977395.977673} features more than one hundred compiler
passes~\cite{LLVMProject2019}.

In the context of the NMODL Framework, we focus on a few optimization passes with very specific
objectives. By taking
advantage of domain-specific and high-level information that is available in the DSL but later lost
in the lower level \emph{C++} or \emph{CUDA} code, we are able to provide additional significant
improvements in code performance. For example, all NMODL \texttt{RANGE}, \texttt{ASSIGNED}, and
\texttt{PARAMETER} variables are translated to \emph{double} type variables in \emph{C++}. Once this
transformation is done, \emph{C/C++} compilers can no longer infer these high-level semantics from these
variables. Another example is \texttt{RANGE} to \texttt{LOCAL} transformations with the help
of \texttt{PROCEDURE} inlining discussed in \autoref{sec:nmodl_dsl}. All \texttt{RANGE} variables in
the NMODL DSL are converted to array variables and are dynamically allocated in \emph{C++}. Once
this transformation is done, the \emph{C/C++} compiler can only do limited optimizations.

To facilitate the DSL level optimizations summarized in \autoref{sec:nmodl_dsl}, we have implemented
the following optimization passes. As most of these optimization passes are also commonly
used in compilers today, we only summarize their role in the context of NMODL DSL optimizations.


{\bf Inlining} : To facilitate optimizations such as \texttt{RANGE} to \texttt{LOCAL} conversion and
facilitate other code transformations, the \emph{Inlining} pass performs code inlining of
\texttt{PROCEDURE} and \texttt{FUNCTION} blocks at their call sites.

{\bf Variable Usage Analysis} : Different variable types such as \texttt{RANGE}, \texttt{GLOBAL},
\texttt{ASSIGNED} can be analysed to check where and how often they are used. The \emph{Variable
Usage Analysis} pass implements \emph{Definition-Use (DU)} chains~\cite{Kennedy1978} to perform data
flow analysis.

{\bf Localiser} : Once function inlining is performed, \emph{DU} chains can be used to decide which
\texttt{RANGE} variables can be converted to \texttt{LOCAL} variables. The \emph{Localiser} pass is
responsible for this optimization.

{\bf Constant Folding and Loop Unrolling} : The \texttt{KINETIC} and \texttt{DERIVATIVE} blocks can
contain coupled ODEs in \texttt{WHILE} or \texttt{FOR} loop statements. In order to analyse these
ODEs with SymPy (see \autoref{sec:ode}), first we need to perform constant folding to know the
iteration space of the loop and then perform \emph{loop unrolling} to make all ODE statements
explicit.

All above-described optimization passes operate on the \emph{AST} and are implemented using the
\emph{Visitor Pattern}.

\subsection{Code Generation}
\label{subsec:codegeneration}

Once DSL and symbolic optimizations (see \autoref{sec:ode}) are performed on the AST, the NMODL Framework is ready to proceed
to the code generation step (cf. \autoref{fig:architecture}). Table~\ref{tab:targets} summarizes
the various target languages supported for different hardware platforms. The \emph{C++} code
generator plays a special role, since it constitutes the base code generator extended by all other
implementations.  This allows easy implementation of a new target by overriding only necessary
constructs of the base code generator.

While our initial code generation target was \emph{C++}, allowing us to validate the code against
the original implementation, our primary target is the \emph{Intel SPMD Program Compiler} (ISPC)
\cite{Pharr2012}. ISPC is built on top of the LLVM compiler framework, leveraging its modular
architecture and powerful code generation backend supporting a multitude of hardware platforms. We chose ISPC for
its performance portability and support for all major vector extensions on x86 (SSE4.2, AVX, AVX2,
AVX-512), ARM NEON and NVIDIA GPUs (using NVPTX) giving us the ability to generate optimized SIMD
code for all major computing platforms. Since the standard library of ISPC is lacking double
precision implementations for several transcendental functions such as \texttt{exp}, we provide our
own implementations based on VDT \cite{Piparo2014}. Measurements of our ISPC
implementation of \texttt{exp} show on-par performance with state-of-the-art mathematical libraries
such as Intel's SVML.

We have, furthermore, extended the \emph{C++} target with an
OpenMP and an OpenACC. These two code generators emit code that is largely identical to the \emph{C++}
code generator but add appropriate pragma annotations to support OpenMP shared-memory parallelism
and OpenACC GPU acceleration. Finally, our code-generation framework supports CUDA as a main backend
to target NVIDIA GPUs. We were able to integrate NMODL as a code generation backend for CoreNEURON
with only a few minor modifications to the build system. The CoreNEURON build process calls NMODL to
generate \emph{C++}/\emph{ISPC}/\emph{OpenACC}/\emph{CUDA} files, which are subsequently compiled
and linked to the simulator binary using the appropriate compilers. We can thus use NMODL as a
drop-in replacement for the current \texttt{mod2c} transpiler, allowing us to easily perform
numerical validation and performance benchmarking.  The \emph{OpenACC} and \emph{CUDA} backend
targeting GPUs require changes in CoreNEURON and this is an ongoing effort.

\begin{table}
  \begin{tabular}{l l l}
    \toprule
    Target Language & Hardware Platform & Features\\ \midrule
    C++               & CPUs              & x86 SIMD using \texttt{ivdep} \\
    C++ + OpenMP        & CPUs              & multithreading \\
    C++ + OpenACC       & CPUs, GPUs        & accelerators \\ \midrule
    ISPC            & x86, ARM          & native SIMD \\
                    & NVIDIA GPGPUs     & native SPMD \\ \midrule
    CUDA            & NVIDIA GPGPUs     & native SPMD \\ \bottomrule
  \end{tabular}
  \caption{Summary of supported target platforms by NMODL Framework. In addition
  to the standard \emph{C++} target, ISPC is used to target various hardware platforms with
  optimal SIMD performance}
  \label{tab:targets}
\end{table}

\subsection{AST to NMODL Transformation}

Various optimization passes discussed in ~\autoref{subsec:optimizationpasses} are performed at the DSL
level by keeping the languange semantics intact as part of the AST. We have implemented a
\emph{AST-to-NMODL} transformation compiler pass to convert optimized AST back to the NMODL DSL form.
This is not only useful to track the AST transformations during development/debugging but also to
generate optimized NMODL DSL that can be consumed by existing transpilers like \emph{nocmodl} and
\emph{mod2c} which lack such optimization capabilities.

\begin{figure*}
  \includegraphics[width=1.0\textwidth]{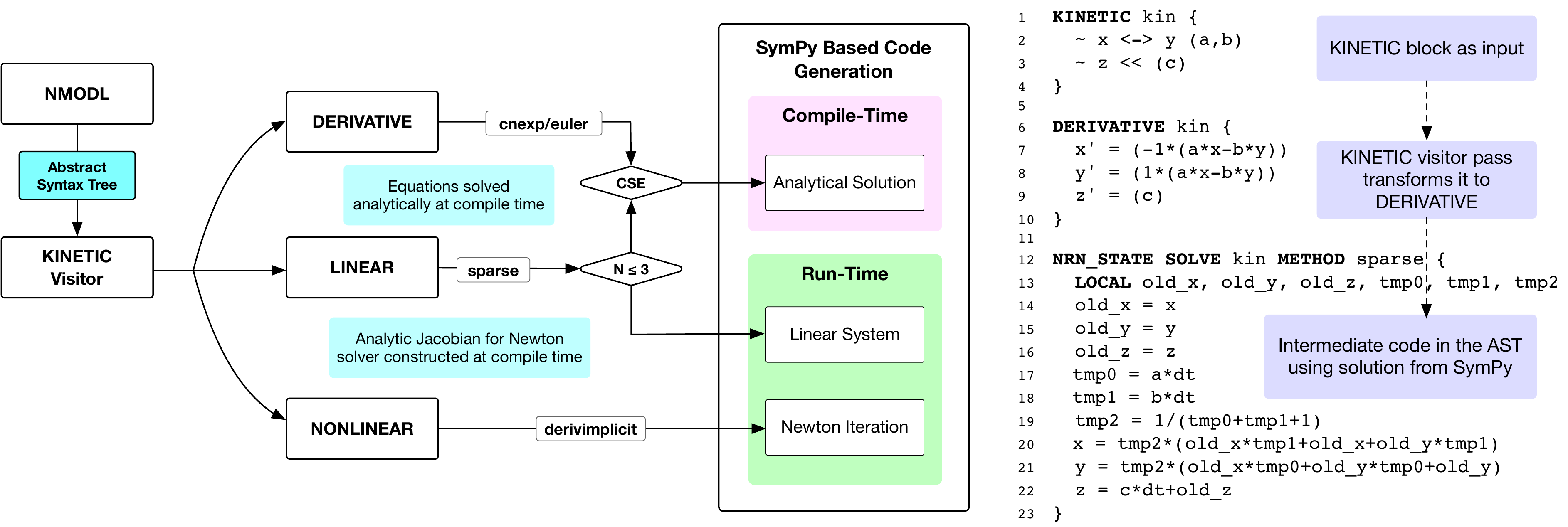}
  \caption{
    On the left, unified ODE solver workflow showing ODEs from different NMODL constructs 
    either produces compile--time analytical solutions, or run--time numerical solutions.
    On the right, example of \texttt{KINETIC} block and its transformation to SymPy based
    solution.
    }
  \label{ode-workflow}
\end{figure*}

\section{ODE Solvers}
\label{sec:ode}

\subsection{Overview}
NMODL allows the user to specify the equations that define the system to be simulated in a variety of ways.
\begin{itemize}
    \item The \texttt{KINETIC} block describes the system using a mass action kinetic scheme of
      reaction equations.
    \item The \texttt{DERIVATIVE} block specifies a system of coupled ODEs (note that
      any kinetic scheme can also be written as an equivalent system of ODEs.)
    \item Users can also specify systems of algebraic equations to be solved. The \texttt{LINEAR} and
      \texttt{NONLINEAR} blocks respectively specify systems of linear and nonlinear algebraic equations
      (applying a numerical integration scheme to a system of ODEs
      typically also results in a system of algebraic equations to solve.)
\end{itemize}

To reduce duplication of functionality for dealing with these related systems of equations, we
implemented a hierarchy of transformations as shown in Figure~\ref{ode-workflow}.
First, any \texttt{KINETIC} blocks of mass action kinetic reaction statements
are translated to \texttt{DERIVATIVE} blocks of the equivalent ODE system. Linear and independent ODEs
are solved analytically. Otherwise a numerical
integration scheme such as implicit Euler is used which results in a system of
algebraic equations equivalent to a \texttt{LINEAR} or \texttt{NONLINEAR} block.
If the system is linear and small, it is solved analytically at compile time using symbolic
Gaussian elimination. Optionally, Common Subexpression Elimination (CSE)~\cite{Cocke1970} can then be applied.

If the system is linear and large, it is solved (at run time) using a lower--upper (LU) matrix
decomposition.  Non-linear systems of equations are solved at run time by Newton iteration, which
makes use of the analytic Jacobian calculated at compile time.

The use of external library solvers for analytic and numerical ODEs offers far
superior efficiency, precision, and simplicity compared to the legacy implementation.
The analytic ODE solver uses SymPy~\cite{Meurer2017}, a Python library for symbolic calculations,
which can simplify, differentiate and integrate symbolic mathematical expressions. Our analytical
solver replaces the purely numerical approach used in other NMODL source-to-source compilers and
simulators. It allows us to perform some computations analytically at compile time that were
previously carried out at run time at each time step using approximate numerical differentiation. This increases
both the numerical accuracy and performance of simulations. The numerical ODE solver uses the
Eigen~\cite{Guennebaud2010} numerical linear algebra \emph{C++} template library, which produces highly
optimized and vectorized routines for solving systems of linear equations.

\subsection{SymPy}
We use the SymPy library to perform algebraic simplifications, differentiate and integrate
expressions, and identify common sub expressions. Performing these operations analytically and
exactly at compile time removes a source of numerical errors and increases the overall
run time performance.

Linear and independent ODEs have been typically replaced by an analytic solution that treats the
coefficients as constant over a time step. NMODL increases the runtime performance by performing
algebraic simplification and optionally replacing computationally expensive exponential calculations
with the (1, 1) \emph{Pade approximant}~\cite{Cambridge2007}, consistent with the overall second order correct simulation accuracy
(as suggested in~\cite{Casalegno2016}, and implemented in~\cite{Akar2019}).

For coupled ODEs, the implicit Euler numerical integration scheme is applied which results in
a set of simultaneous algebraic equations. For a linear systems of equations,
such as those that arise from a set of linear reaction equations in a \texttt{KINETIC} block,
the \texttt{sparse} solver method is used. For non-linear systems, the \texttt{derivimplicit} solver method is used.

The \texttt{sparse} solver chooses from two solution methods, depending on the size of the system to
be solved.  For small systems (three or less equations), the system is solved by symbolic Gaussian
elimination at compile time. Additionally the solver
will also optionally perform CSE. Figure~\ref{ode-workflow} contains a simple example of this
process.  For larger systems, performing Gaussian elimination at compile time may not result in a
numerically stable solution.  Instead we construct a linear system of equations, which can be solved
numerically using Eigen.

The \texttt{derivimplicit} solver constructs a system of non--linear equations, which we solve using
Newton's method at run time. We therefore compute the system's Jacobian, which is then used in the
iterative solver.  The existing solver in NEURON calculates the Jacobian by numerical approximation at
run time. Using Sympy we are, however, able to analytically differentiate the system of non--linear
equations to construct the exact Jacobian at compile time.  By using the exact Jacobian we are able
to reduce the number of Newton solver iterations required, as well as reduce the time per iteration.

A further improvement which makes use of SymPy involves the conductance, which is the derivative
with respect to voltage of a current. By default in NEURON these derivatives are numerically
approximated at run time. This introduces an additional numerical error and generally doubles
the computation time.
To avoid this issue, the \texttt{CONDUCTANCE} keyword was previously introduced in
NMODL~\cite{Kumbhar2016}, which allows the user to manually indicate the presence of an
\texttt{ASSIGNED} variable that represents the current derivative.
Since the AST can be analyzed to determine that a conductance calculation needs to be performed, our implementation
automatically generate an appropriate local conductance variable and \texttt{CONDUCTANCE} statement
to further reduce unnecessary run time computation.

\subsection{Eigen}

To solve linear systems of equations we use the Eigen \texttt{PartialPivLU()} routine~\cite{Jacob}, which
performs LU decomposition with partial pivoting, a fast and stable algorithm for square invertible
matrices.

For non--linear systems, we offer a simple implementation of Newton's method,
which is an iterative solver that requires a linear system to be solved at each iteration.
For large system matrices this linear solve also uses \texttt{PartialPivLU()}.
For $4\times4$ or smaller matrices, however, we directly invert the matrix using the
\texttt{invert()} routine.
For such small matrices this is numerically safe.
Additionally this method does not involve any pivoting and hence does not incur any branching,
which leads to significantly faster code than LU decompositions.

In general, using Eigen routines in favor of our own implementations provides two main benefits.
First, the code becomes more maintainable: often a long and complex routine can be replaced with a single
Eigen call, which renders the code easier to read and maintain. Second, the code performance is
improved: Eigen offers highly tuned and parameterized routines and specialized implementations for different
matrix sizes through generic programming. Replicating or beating the performance offered by a
specialized numerical library is hardly practical and beyond the aim of this work.

\begin{figure*}
  \includegraphics[width=1.0\textwidth]{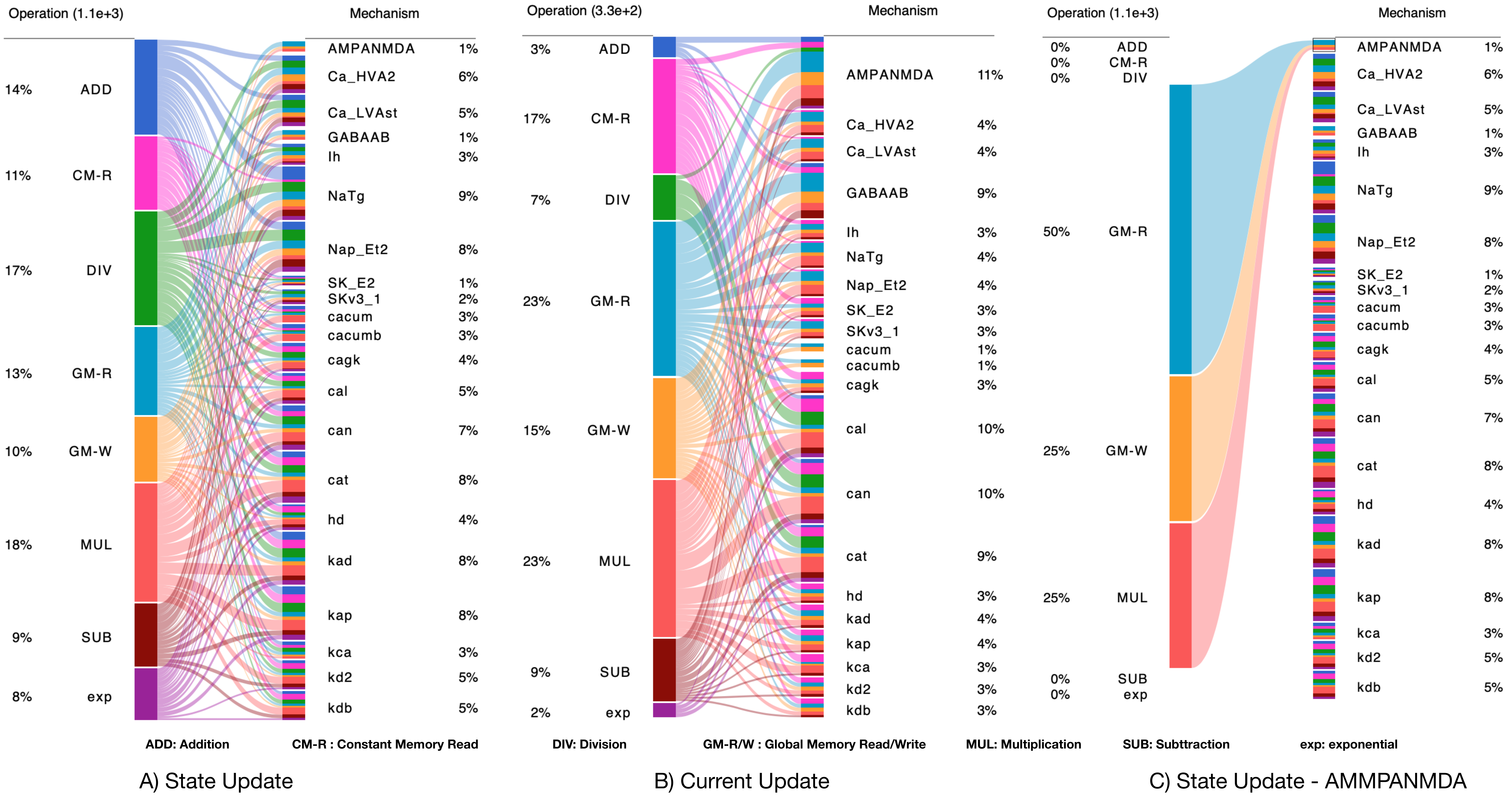}
  \caption{Performance characterization of different mechanisms using high level
  AST analysis from NMODL Framework : A) and B) shows global view of different operations performed
  by \emph{State Update} and \emph{Current Update} kernels respectively and in C) for AMMPANMDA one
can zoom into a specific mechanism to understand detailed performance characteristics}
  \label{fig:perf-bigraph}
\end{figure*}

\begin{figure}
  \includegraphics[width=\linewidth]{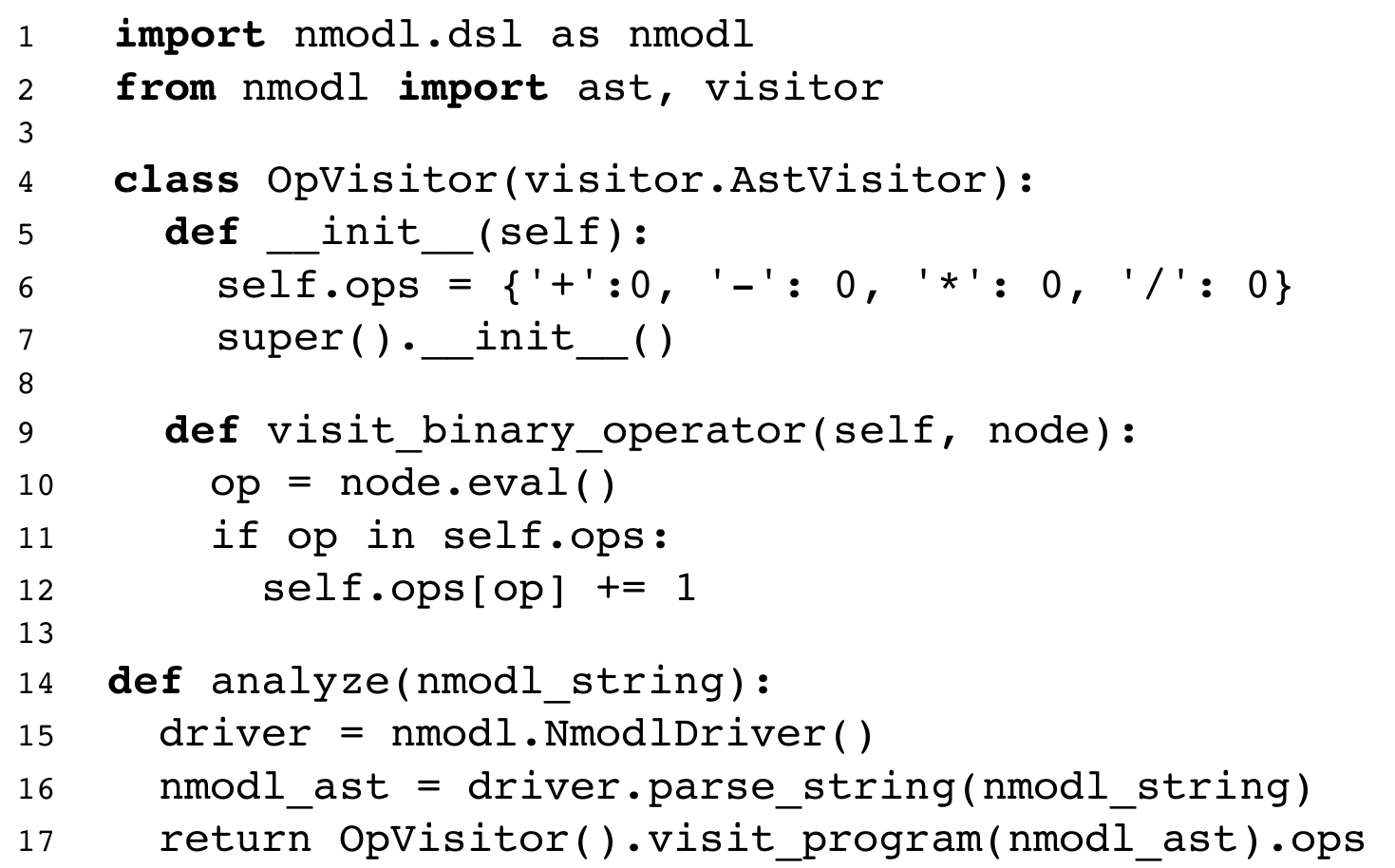}
  \caption{A simple example of a user-defined AST visitor to count FLOPs showing how C++ visitors can be extended using Python API}
  \label{lst:pyast} 
\end{figure}

\section{Analysis Capabilities}
\label{sec:analysis}

An important goal of the NMODL Framework is to not only generate fast code but to also provide the
computational neuroscientist with a versatile and easy-to-use tool for introspecting and
programmatically modifying models written in NMODL DSL.  Currently, the NEURON NMODL database counts
over 6,300 mechanism descriptions contributed by a world wide community of researchers. Until now, any
meta-analysis of these NMODL files had to be done either manually or using home-made tools. The
problem with such tools is that their parsing capabilities are often fragile since they rely only on
regular expressions or simple manually implemented parsers. In contrast, the NMODL Framework
features a robust parser along with a complete intermediate representation and a powerful visitor
framework.  Using the AST and visitor classes, a programmer can do any analysis or modification they 
wish. The downside is that using the framework requires knowledge of modern C++ and ideally some
experience with compiler frameworks. We therefore provide Python bindings to all AST classes and the
visitor framework using the pybind11 library \cite{Jakob2017}. Additionally, we expose the ODE
solver tools described in section~\ref{sec:ode} in the Python API. This gives the user access to all
the ODE solving and simplification functionality available within the framework. When combined with SymPy, this yields
a comprehensive analysis framework for mechanism dynamics described in the NMODL DSL.

Figure~\ref{lst:pyast} shows an example of a simple AST visitor that keeps track of binary
arithmetic operator counts. The programm is self explanatory and shows how the C++ AST and visitor
classes are exposed in Python and can not only be instantiated, but also subclassed and extended.
Using these capabilities, we can build rich analysis tools to introspect various aspects of the
models. For example, \autoref{fig:perf-bigraph} shows a Bipartite graph with theoretical performance
characteristics of different mechanisms from benchmarks discussed in~\autoref{sec:benchmark}.
\autoref{fig:perf-bigraph}A shows that the \emph{State Update} kernels have more high latency
operations like \emph{division} and \emph{exponential} (compute bound) compared to the more memory
and bandwidth bound \emph{Current Update} kernels of~\autoref{fig:perf-bigraph}B. This has been
previously shown with hardware counter analysis in other studies~\cite{Cremonesi2019,Kumbhar2016}.  The
detailed validation of these theoretical performance metrics is beyond the scope of this paper but,
as discussed in ~\autoref{sec:results}, these high level performance characteristics can be directly
correlated with the observed performance.

\section{Benchmarks}
\label{sec:benchmark}

\begin{figure*}
  \includegraphics[width=1.0\textwidth]{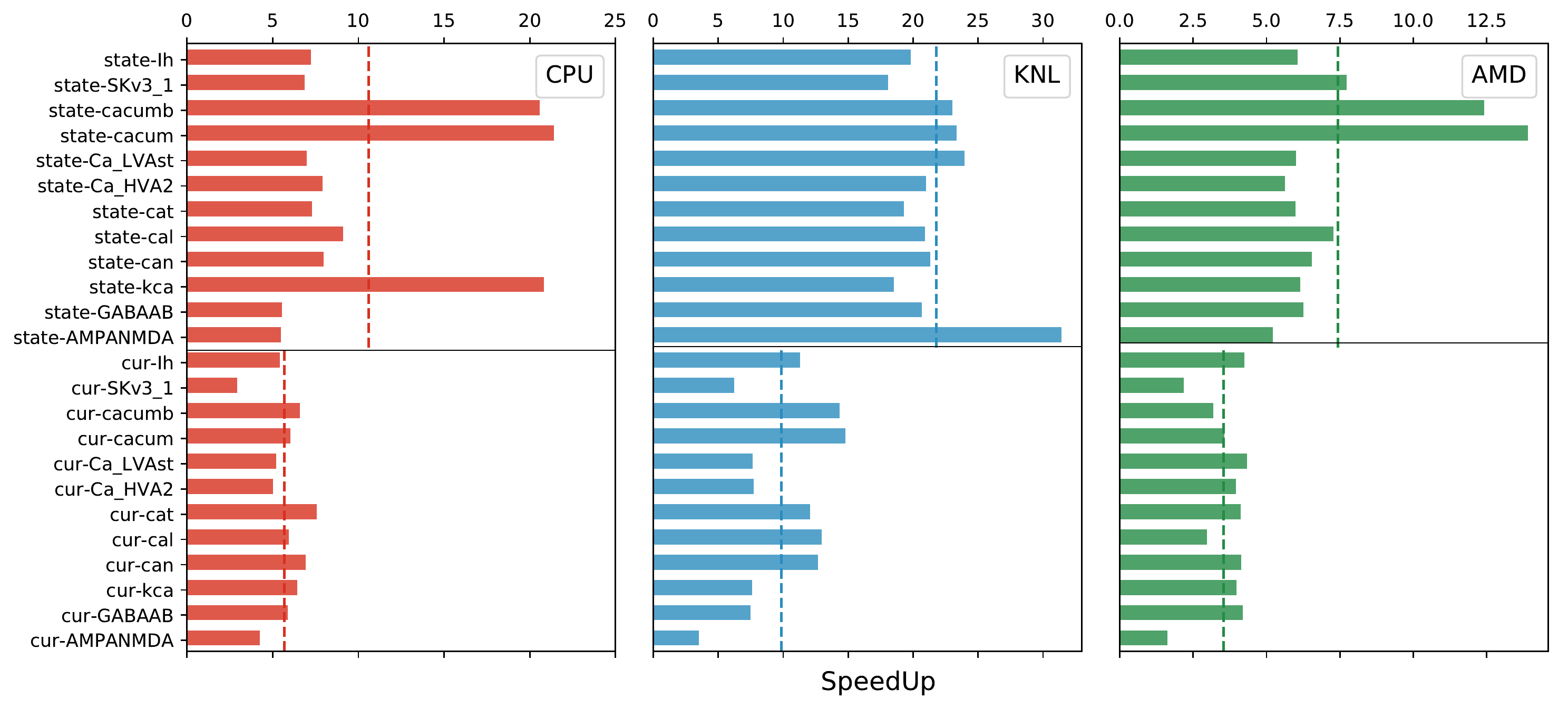}
  \caption{Speedup of different channels selected from neocortex and hippocampus circuits on Intel Skylake, Intel KNL and AMD Naples platform}
  \label{fig:channel_bench}
\end{figure*}

To evaluate the achieved performance gains through NMODL, we have performed
comprehensive benchmarks on three major production hardware platforms
summarised in Table\ref{table:benchmark-systems}. Since in this work we
are primarily interested in the on-node performance increase from the generated code,
we focus on single node benchmarks. We utilize all cores by pinning one MPI process on each core. 
From previous work on the parallel scalability
of NEURON and CoreNEURON, those single node improvements will translate directly
into equivalent improvements for parallel execution.

Benchmarks performed on the Intel platforms were compiled
with Intel Parallel Studio 2018.1, while those on the AMD platform were compiled with GCC 7.3.0. 
All benchmarks have been compiled with \texttt{-O2 -xHost} and \texttt{-O2} flags respectively.
Note that these benchmarks are not meant to compare one
platform with other and hence we did not investigate the full spectrum of hardware
capabilities on each platform (e.g. \emph{hyperthreading}).

\begin{table}
\footnotesize
\begin{tabular}{ll}

    \toprule
    Intel Skylake & \\
    Processor     & 2 Xeon 6140, 36 cores @2.3GHz, 384GB DRAM \\
    Toolchain     & Intel 2018.1 and HPE-MPI (MPT) \\

    \midrule
    Intel KNL     & \\
    Processor     & Xeon Phi (7230), 64 cores @1.3GHz, 16GB MCDRAM, 96GB DRAM  \\
    Toolchain     & Intel 2018.1 and HPE-MPI (MPT) \\

    \midrule
    AMD Naples    & \\
    Processor     & 2 AMD EPYC 7451, 24 cores @2.3GHz, 128GB DRAM  \\
    Toolchain     & GCC 7.3.0, Intel MPI 2019\\

    \bottomrule

\end{tabular}
\caption{Details of Benchmarking Systems}
\label{table:benchmark-systems}
\end{table}
We selected two brain tissue models: a somatosensory microcircuit and a hippocampus region model.
The first, a somatosensory cortex microcircuit of a young rat published by the Blue Brain Project has 55 layer-specific
morphological types and 207 morpho-electrical types~\cite{Markram2015}. The second, a model of a rat
hippocampus CA1~\cite{HumanBrainProject} is built as part of the European Human Brain Project and has 13
morphological types and 17 morpho-electrical types. These models are selected because they are
computationally expensive and have a large number of mechanisms which allow us to assess
performance benefits for different types of kernels used in production simulations. Based on these
two models, we presented results for two benchmarks:

{\bf Channel Benchmark} : This benchmark consists of 21 different morpho-electrical types selected to include all the
unique mechanisms from somatosensory cortex and hippocampus CA1 models. A total of 2,304 cells are
created without network connectivity as this benchmark is designed to measure performance of code
generation for individual mechanisms.

{\bf Simulation Benchmark} : This benchmark measures the overall performance improvement for production
simulations. We used subcircuits of 1,000 cells from the two models above simulating one second of
biological time using a timestep of \emph{0.025 ms}.

\begin{figure*}
  \includegraphics[width=1.0\textwidth]{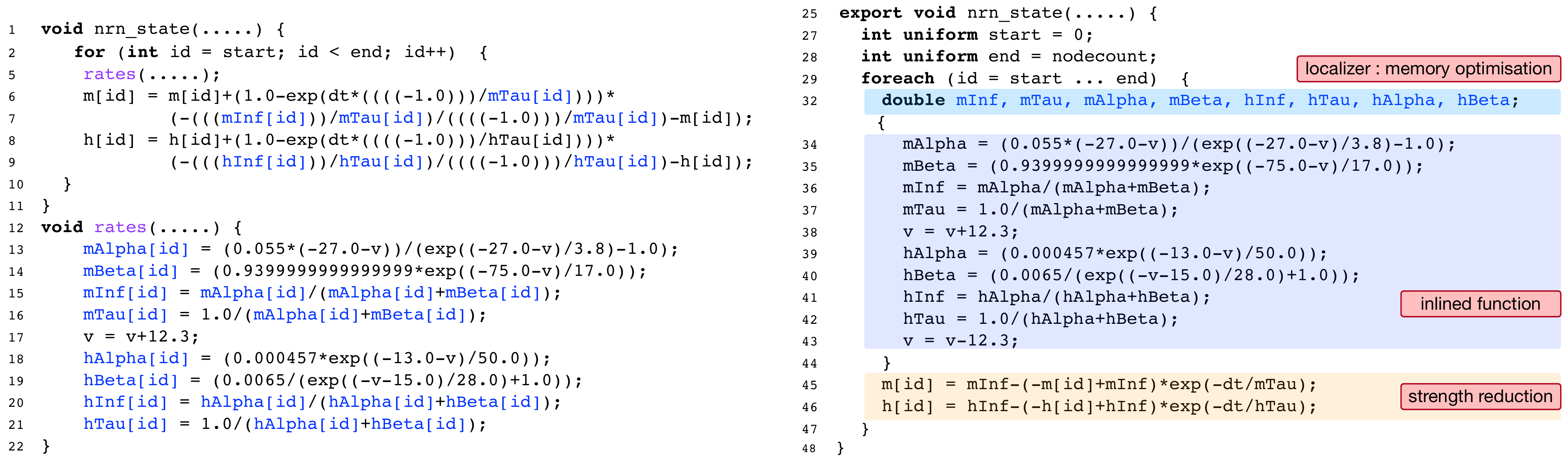}
  \caption{Comparison of unoptimized and optimized code (from calcium mechanism, simplified):
  On the left, \texttt{nrn\_state} kernel generated from \texttt{DERIVATIVE} block calling
  \texttt{rates} function as generated by \emph{nocmodl} and \emph{mod2c}. On the right,
  corresponding ISPC code generated by the NMODL Framework after the DSL optimizations
  and SymPy transformations described in \autoref{subsec:optimizationpasses} and \autoref{sec:ode}.}
  \label{fig:opt_code_ex}
\end{figure*}

\section{Results}
\label{sec:results}

In order to understand the impact of the various optimizations and performance of generated code, we
implemented a first draft of code generation backend for the CoreNEURON library. The same code
generation backend can be adapted for the original NEURON simulator in the future. As the CoreNEURON
library is a compute engine executed under the NEURON simulator producing binary identical results,
the performance results can be directly compared between NEURON and CoreNEURON. More details about
CoreNEURON library can be found in Kumbhar et al., 2010~\cite{Kumbhar2019}.

We have performed benchmarks for a representative set of NMODL mechanisms as described in
\autoref{sec:benchmark}.

For calculating the speedup, we compare the runtimes of the mechanisms translated using NMODL
Framework with ISPC backend and simulated using CoreNEURON with the same mechanisms compiled with
the NEURON simulator using \emph{nocmodl}. The results of these benchmarks are shown in
\autoref{fig:channel_bench}. We restrict ourselves here to the 12 most expensive channels and
highlight timing for the two main computational parts: the \emph{State Update} (denoted as
\emph{state-channel-name}) and the \emph{Current Update} (denoted as \emph{cur-channel-name}).  The
three columns correspond to the three tested hardware platforms. Finally, we show as dotted lines
the average speedups achieved for the shown channels. Generally we observe a higher speedup on
\emph{State Update} kernels than on \emph{Current Update} kernels (e.g. on Intel Skylake $9.6\times$ vs.  $5.5\times$).
This is due to the fact that \emph{State Update} kernels are typically computationally more
expensive, with a higher FLOP per byte ratio than for \emph{Current Update} kernels (see ~\autoref{sec:analysis}).
When comparing the three different benchmark platforms, we notice that on KNL, and particularly for
\emph{State Update} kernels, the best results are achieved. We attribute this to the rather poor
performance of the \emph{nocmodl} code generation backend with NEURON instead of particularly
higher-than-expected performance of the here-presented framework. Finally, we notice that especially
in the \emph{State Update} kernel the availability of AVX-512 vector units, with optimal memory
layout offers a performance advantage as can be seen in the higher performance of the two Intel
platforms compared with the AMD EPYC platform, which only offers AVX2.

When looking at top performers we notice that several of the high-level optimizations
described in sections \ref{subsec:optimizationpasses} and \ref{sec:ode} are at least equally if not more important
than the generation of vectorized code. We observe that particularly our optimizations on the ODE
statements using SymPy based solvers (e.g. \emph{state-cacum}) can lead to speedups of more than $20\times$ ($12\times$ on AMD EPYC).

In \autoref{fig:opt_code_ex} we present an example \emph{State Update} kernel generated by the NMODL Framework.
For brevity, we have removed some of the unnecessary variables and initializations. \emph{nocmodl}
and \emph{mod2c} generate code equivalent to the one shown on the left but with more complex data
structures and an indirect addressing scheme. In practice we have observed that auto-vectorization for these kernels
is highly dependent on the compiler and hardware platform and cannot be relied upon.
On the right side we show the generated ISPC SPMD code.  The code is
compiled in a separate object file and kernel routines must be denoted with \texttt{export} to be
callable from the corresponding \emph{C} wrappers. Furthermore, we note the \texttt{foreach}
keyword available in ISPC, allowing the loop body to be executed in so-called gangs of
vector-elements. At a higher level this example illustrates the result of three NMODL Framework
optimization passes. First, the \texttt{rates} function has been inlined.
Second, the localizer pass transforms global variables to intermediate local variables,
drastically reducing memory requirements and memory bandwidth. Third,
the sympy pass provides analytic integration further
reducing the computational cost of the kernel. Several channels from benchmarks (e.g. \emph{CaT}) enjoy these
optimizations and show speedups of more than $6\times$ on all benchmark platforms.

\begin{table*}[h]\centering
\small
\begin{tabular}{llcccccc}
\toprule 
\multirow{2}{*}{Simulation} & \multirow{2}{*}{Component} & \multicolumn{3}{c}{Intel Skylake} & \multicolumn{3}{c}{Intel KNL} \\
\cmidrule(lr){3-5} \cmidrule(lr){6-8}
& & NRN-NOCMODL & CN-MOD2C & CN-NMODL & NRN-NOCMODL & CN-MOD2C & CN-NMODL \\
\midrule
\multirow{4}{*}{Hippocampus} & State Update & 1298.83 & 316.11 & 138.15 & 4786.48 & 713.7 & 233.64 \\
  & Current Update & 1102.56 & 239.84 & 165.53 & 2732.88 & 181.94 & \cellcolor{red!25}348.72 \\
  & Other & 154.13 & 44.08 & 40.78 & 828.94 & 236.11 & 203.24 \\
  & Total & 2555.52 & 600.04 & 344.46 & 8348.29 & 1131.74 & 785.6 \\
  & Speedup wrt NEURON & 1 & 4.26 & 7.42 & 1 & 7.38 & 10.63 \\
\midrule
\multirow{4}{*}{Plasticity} & State Update & 199.8 & 33.17 & 25.41 & 846.5 & 63.77 & 43.86 \\
  & Current Update & 179.93 & 36.2 & 21.58 & 422.39 & 38.43 & \cellcolor{red!25}61.73 \\
  & Other & 47.71 & 19.1 & 17.12 & 310.93 & 89.43 & 78.49 \\
  & Total & 427.44 & 88.46 & 64.11 & 1579.82 & 191.63 & 184.08 \\
  & Speedup wrt NEURON & 1 & 4.83 & 6.67 & 1 & 8.24 & 8.58 \\
\bottomrule
\end{tabular}
\caption{Absolute time(s) and speedup of the hippocampus and somatosensory cortex simulations on Intel Skylake and Intel
    KNL platform using NEURON with \emph{nocmodl} (NRN-NOCMODL), CoreNEURON with MOD2C (CN-MOD2C) and CoreNEURON with
    NMODL Framework (CN-NMODL)}
\label{table:absolute-time-speedup}
\end{table*}

\begin{figure}
  \includegraphics[width=0.5\textwidth]{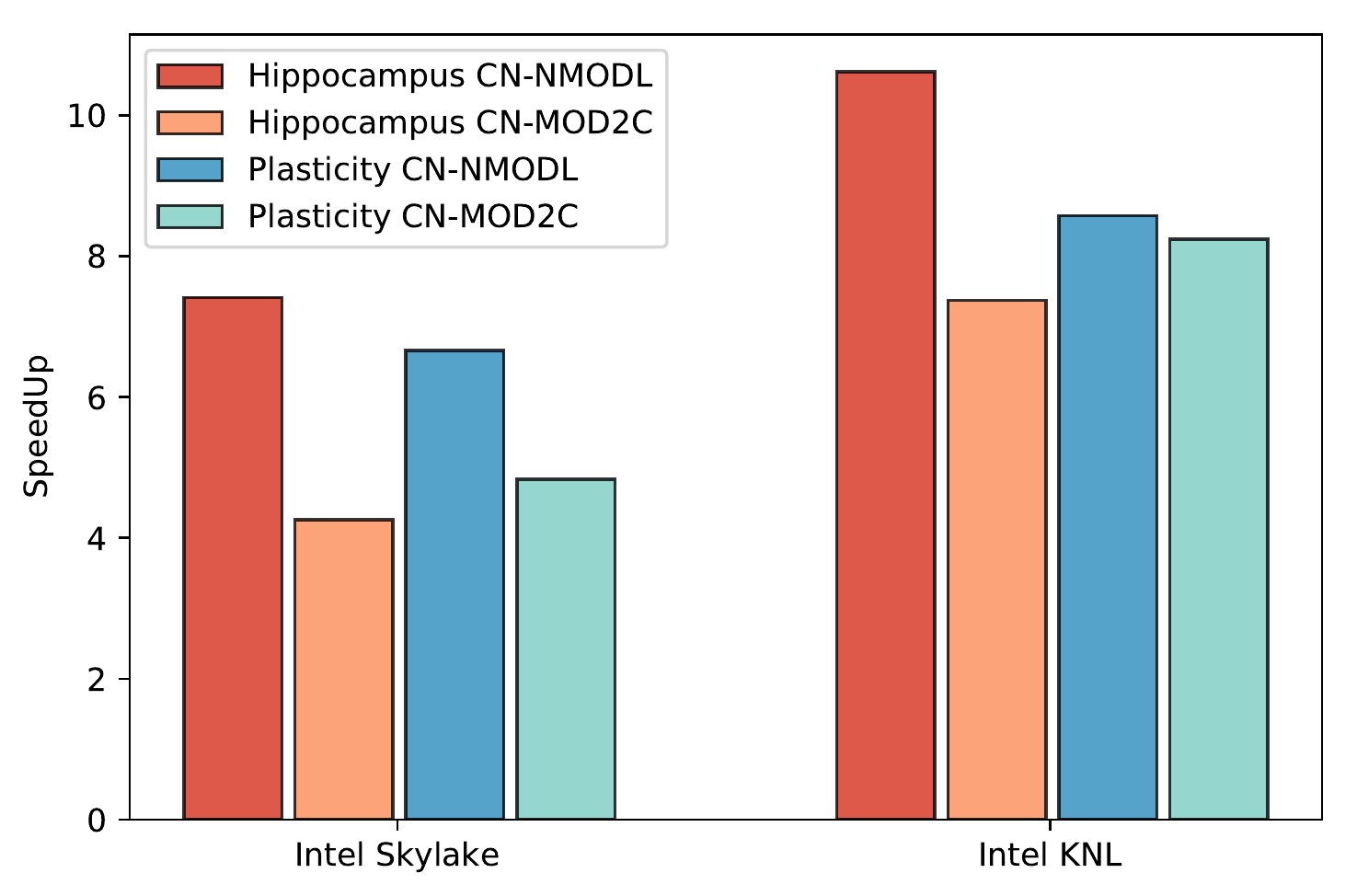}
    \caption{Speedup of \emph{CN-MOD2C} and \emph{CN-NMODL} compared to \emph{NRN-NOCMODL}
    for somatosensory cortex and hippocampus CA1 models.}
  \label{fig:sim_speedup}
\end{figure}

\autoref{fig:sim_speedup} shows the speedup achieved for whole simulations for the hippocampus CA1 and
somatosensory cortex models. We compared the performance with three different configurations. The first
configuration (\emph{NRN-NOCMODL}) uses the \emph{NEURON} simulator with \emph{nocmodl} as the code
generation backend. The second configuration (\emph{CN-MOD2C}) uses the \emph{CoreNEURON} library with
\emph{MOD2C} as a code generation backend. The third configuration (\emph{CN-NMODL}) uses the
\emph{CoreNEURON} library with the here-presented NMODL Framework as a code generation backend.

The NMODL Framework shows up to $7\times$ speedup on Skylake and up to $11\times$ speedup on the KNL
platform. The hippocampus model shows a larger speedup compared to the somatosensory cortex model because
it uses \emph{cacum}, \emph{cacumb} and \emph{kca} mechanisms with the \emph{derivimplicit}
integration scheme. The Eigen based solver implementation in NMODL brings additional performance improvements. When
compared with \emph{CN-MOD2C}, \emph{CN-NMODL} shows up to $2\times$ performance improvement. Note
that the \emph{CN-MOD2C} is heavily dependent on auto-vectorization capabilities of the compiler. For example,
if the GCC compiler is used instead of Intel, \emph{CN-NMODL} becomes $5\times$ faster compared to
\emph{CN-MOD2C}. The breakdown of the
execution times is shown in \autoref{table:absolute-time-speedup}. As discussed before,
\emph{State Update} and \emph{Current Update} represent the time spent in the execution of code
generated from NMODL DSL and the rest of the time is shown as \emph{Other}. On the Intel KNL
platform the \emph{Current Update} kernel is $\sim$$2\times$ slower in \emph{CN-NMODL} comapred to \emph{CN-MOD2C}.
This is due to a performance issue found with ISPC when atomic reductions are used in the kernel.
This performance issue will be addressed in a future release of NMODL Framework.

\section{Conclusions}

Having real-world scientific applications make efficient use of modern computer architectures' performance features 
is an involved task and heavily relies on the successful combination of using optimized libraries, auto-vectorizing
compilers, and exposure of parallelism and hints by the programmer. Good performance on one architecture
does not imply automatically good performance on the next architecture. In times of increasing architectural
diversity this poses real challenges.

Additionally, many scientific applications do not encode only a single mathematical problem, but the
scientific users provide the mathematical equations that need to be integrated by the solvers on a
case by case basis. In the worst case, this can severely impact the success of auto-vectorization.
In the best case, the way the users express their specific equations, e.g. through DSLs may help in producing optimized code.

In this paper we presented a novel NMODL Code Generation Framework for the DSL of the widely used
NEURON simulator. The DSL constructs are translated into an AST that lends itself to specific optimization 
passes before it is handed off to different backends for generation of optimized code for the target platform. 

We have implemented optimization passes that relate to straight-forward transformation of the DSL code, but also 
more advanced optimization passes that intercept ODE statements for which an analytical solution can be used instead
of having to resort to numerical integration. This functionality is built on top of the SymPy and Eigen libraries. 

For code generation we have developed backends for C++ and OpenMP targeting CPUs and ISPC to
target a wide variety of CPU architectures providing optimal SIMD performance and reducing the dependency on
auto-vectorization capabilities of the compiler. Furthermore, we have developed both a CUDA backend
specifically with NVidia GPUs in mind as well as a more generic OpenACC. Both GPU backends will,
however, require more integration work with the simulators and benchmarking.

We have benchmarked kernels from production simulations of two different large-scale brain tissue models on Intel SKL, Intel KNL and 
AMD EPYC platforms. On those individual kernels, we saw performance improvements from $5\times$ to
$20\times$. In order to test
how those kernel improvements translate into speedup of the entire simulations (which use the kernels in different ratios or not at all),
we benchmarked production simulations on Intel KNL and Intel SKL platforms. Compared to the regular
NEURON simulation environment, a speedup of $6-10\times$ has been observed. Compared to an optimized version of
the NEURON simulator, CoreNEURON, which heavily relies on auto-vectorization of the compiler, the work presented
here nonetheless resulted in a speedup of up to $2\times$.

Beyond the absolute performance, a central goal of our effort was the ability to parse all
previously published models. By using the grammar specification from the original NEURON NMODL
language, we were able to demonstrate compatibility with 6,370 channels from the public model
repository ModelDB. We furthermore took care to maintain the language semantics of the DSL in the
AST, allowing retranslation of the AST optimization into the DSL constructs, thus making the
optimizations available to the regular NEURON simulator without having to rely on our code
generation backends (albeit with reduced overall speedup). Lastly, our framework is exposed through
a Python interface, providing great flexibility to use NMODL as a generic NMODL parsing framework
and build new tools on top of it.

{\bf Availability} \newline NMODL Framework is open sourced and available on GitHub\cite{Kumbhar}.

%
\begin{acks}
This work has been funded by the EPFL Blue Brain Project (funded by the Swiss ETH board),
NIH grant number R01NS11613 (Yale University) and partially funded by the European Union's
Horizon 2020 Framework Programme for Research and Innovation under Grant Agreement number
785907 (Human Brain Project SGA2). We would like to thank Antonio Bellotta, Francesco Cremonesi,
Ioannis Magkanaris, Matthias Wolf, Samuel Melchior and Tristan Carel for fruitful discussions and 
their contributions to the NMODL development. The AMD system for benchmarking was provided by 
Erlangen Regional Computing Center (RRZE).
\end{acks}
%

\Urlmuskip=0mu plus 1mu\relax
\bibliographystyle{ieeetr}
\bibliography{nmodl}

\end{document}